\documentclass[aps,superscriptaddress,twocolumn,floatfix]{revtex4-2}

\usepackage{graphicx}
\usepackage{amsmath}
\usepackage{amssymb}
\usepackage{booktabs}
\usepackage{xurl}
\usepackage[colorlinks=true,linkcolor=blue,citecolor=blue,urlcolor=blue]{hyperref}

\newcommand{\LEmSR}{LE-$\mu$SR}
\newcommand{\bNMR}{$\beta$-NMR}
\newcommand{\trimspnl}{TRIM.SP-NL}
\newcommand{\workbench}{TRIMSP-NL Workbench}
\newcommand{\ranlux}{RANLUX}
\newcommand{\pcg}{PCG32}
\newcommand{\xoshiro}{\texttt{xoshiro256**}}

\newcommand{\safeincludegraphics}[2][]{%
  \IfFileExists{#2}{\includegraphics[#1]{#2}}{%
    \fbox{\parbox[c][4.0cm][c]{0.90\linewidth}{\centering
      Missing figure file:\\[0.5ex]\texttt{\detokenize{#2}}}}}}

\hypersetup{
  pdftitle={Modernization and Statistical Validation of a Multilayer TRIM.SP Code},
  pdfauthor={Z. Salman and T. Prokscha},
  pdfsubject={Monte Carlo ion implantation software},
  pdfkeywords={TRIM.SP, low-energy muons, ion implantation, Monte Carlo, random-number generator}
}

\begin{document}

\title{Modernization and Statistical Validation of a Multilayer TRIM.SP Code for Low-Energy Muon and Ion Implantation}

\author{Zaher~Salman}
\email[Corresponding author: ]{zaher.salman@psi.ch}
\affiliation{PSI Center for Neutron and Muon Sciences, 5232 Villigen PSI, Switzerland}
\author{Ryan~M.~L.~McFadden}
\affiliation{TRIUMF, 4004 Wesbrook Mall, Vancouver, BC V6T 2A3, Canada}
\affiliation{Department of Physics and Astronomy, University of Victoria,
3800 Finnerty Road, Victoria, BC V8P 5C2, Canada}
\author{Thomas~Prokscha}
\affiliation{PSI Center for Neutron and Muon Sciences, 5232 Villigen PSI, Switzerland}

\date{August 2, 2026}

\begin{abstract}
We describe the staged modernization of a local multilayer TRIM.SP implementation, denoted \trimspnl. The work included behavior-preserving refactoring, correction of localized bookkeeping and initialization defects, support for more than five elements per layer, reproducible build modes, and a runtime-selectable random-number-generator (RNG) interface. The \ranlux\ RNG remains the default reference backend, while \pcg\ and \xoshiro\ provide faster alternatives without changing the input-deck format. The code validation combined exact regression testing with statistical comparisons over 13 representative target multilayer configurations, comprising 143 energy-configuration points per backend and 429 simulations in total. Across the test suite, \pcg\ and \xoshiro\ reduced runtime by factors of 1.98 and 2.01, respectively. The implanted, backscattered, and transmitted fractions, as well as the mean implantation depths, range straggling, and representative depth profiles showed no systematic RNG dependence. Together with a Node.js-based graphical interface for setup, scans, and result collection, the Fortran engine forms the open-source \trimspnl\ Workbench.

\end{abstract}

\maketitle

\section{Introduction}

Monte Carlo simulations of ion transport in condensed matter are used to predict implantation profiles, backscattering and transmission probabilities, sputtering yields, recoil production, and depth-resolved energy deposition. These observables are determined by the combined effects of elastic nuclear collisions and inelastic electronic stopping and are therefore obtained statistically from large ensembles of projectile histories~\cite{Eckstein1991}.

Transport of ions in matter (TRIM) programs implement this transport within the binary-collision approximation. In brief, a projectile is propagated through a sequence of independent two-body encounters with target atoms, with electronic energy loss applied between collisions. The sampled collision partner, impact parameter, and azimuthal angle determine the scattering, along with the transfer of energy and momentum. Recoiling target atoms may themselves be followed if the transferred energy is sufficiently large. TRIM.SP extends this framework with a detailed treatment of surface escape and sputtering-related quantities~\cite{Biersack1984APA} (SP for sputtering).

The same transport machinery is useful in low-energy muon spin spectroscopy (\LEmSR), even when sputtering is not the primary observable. Implantation profiles and backscattering fractions are required to relate a selected muon implantation energy to the depth region sampled in a thin film or multilayer~\cite{Morenzoni2002NIMPRSB}. Related calculations are also useful for light radioactive probes such as $^{8}$Li in beta-detected nuclear magnetic resonance (\bNMR) experiments~\cite{Salman2007NL,Salman2007PRL}.

The code considered here is not the historical TRIM.SP distribution in its original form; it is a local version that was previously extended from a fixed small number of layers to a generalized multilayer geometry. To distinguish this extension from the original program, we refer to the Fortran transport engine as \trimspnl, where ``NL'' denotes a generalized number of layers. The name is introduced here for this implementation and is not intended to imply a separate historical TRIM.SP distribution.

In the \LEmSR\ and \bNMR\ workflow, calculations are commonly performed as scans over implantation energy, incidence angle, layer thickness, and material stack. Aided by a custom graphical user interface (GUI), the user generates the individual input decks, invokes the Fortran executable, and collects the resulting profiles and summary quantities. The same GUI code can operate as a web application \cite{TRIMSPNLWeb} or as a standalone Node.js~\cite{nodejs} application. In standalone use, independent Fortran calculations can be dispatched concurrently because they require no communication with one another.

The legacy Fortran source retained a large shared program scope, numerical branch labels, packed arrays, and monolithic input and output sections. A complete rewrite would have made it difficult to distinguish intended changes from accidental modifications of the transport model. The modernization was therefore incremental. Its objectives were to preserve the established transport physics, make the implementation easier to inspect and test, correct clearly identifiable bookkeeping defects, extend the target description, and reduce the cost of parameter scans by providing optionally faster random-number-generator (RNG) backends.

This paper describes the resulting software structure and its validation. Behavior-preserving changes were tested by exact comparison of the \ranlux-based output, apart from timestamps and timing fields. Changes to the RNG necessarily generate different particle histories and were therefore assessed using statistical comparisons of integral observables and implantation-profile shapes. This validation addresses the consistency of the modernized implementation and RNG backends.

\section{Software architecture and modernization}

\subsection{Simulation workflow}

The Fortran executable performs one Monte Carlo calculation for one input deck. The GUI supplies the higher-level workflow: it defines the target and projectile, creates one or more input decks, launches the executable, and parses the output files. Energy, angle, thickness, and material scans are thus implemented as collections of independent simulations rather than as an additional loop inside the transport engine. This separation keeps the scientific kernel independent of the web or standalone user interface and permits process-level parallel execution on a local workstation.

During the modernization, ChatGPT (OpenAI) was used as an assistive tool to review portions of the legacy Fortran source and to suggest or implement selected refactoring and code changes. All modifications were reviewed by the authors and subjected to the regression and validation procedures described below.

The complete package is referred to as the \workbench. In this terminology, \trimspnl\ denotes only the Fortran transport engine, whereas \workbench\ denotes the engine together with the GUI and scan-management workflow.

\subsection{Behavior-preserving refactoring and build modes}

The first modernization stage exposed the structure of the program without changing its execution. Section banners and a variable dictionary were added to document the major phases of a calculation and the most important arrays, counters, dimensions, and physical quantities. Selected numerical input options were replaced by named constants. These changes improved readability while leaving the numerical path unchanged.

The GUI-generated input reader was then moved into a dedicated subroutine without changing the established input-file layout. The large output blocks were similarly separated into writer routines for the input and target summaries, integral results, implantation profiles, reflection and transmission summaries, sputtering quantities, and matrix-style output tables. The resulting separation also made the Fortran code easier to profile. Profiling showed that the output routines contribute negligibly to the total runtime, so their extraction improved maintainability without a measurable performance penalty.

The build system was divided into release, debug, profiling, and static-link modes. The debug configuration enables runtime checks and backtraces, while the profiling configuration supports tools such as \texttt{gprof}~\cite{Graham1982gprof}. Explicit build modes reduce the risk of comparing executables produced with incompatible compiler options and make regression and timing measurements easier to reproduce.

\subsection{Bookkeeping corrections and target extensions}

Inspection and regression testing exposed several localized bookkeeping and initialization defects. The corrections were restricted to cases in which the intended index, counter, or accumulator was clear from the surrounding implementation. Table~\ref{tab:corrections} summarizes the affected areas. Some defects were latent in the present muon and $^{8}$Li implantation tests because the corresponding branches were rarely active or contributed only to specialized summaries. They were nevertheless corrected to improve robustness for a wider range of input decks.
\begin{table*}[tbh]
\caption{Localized corrections made during the code modernization of \trimspnl.}
\label{tab:corrections}
\centering
\begin{tabular}{ll}
\toprule
Area and correction &
Potentially affected output \\
\midrule
\textbf{Depth bins.} Corrected depth-bin index handling. &
Implantation and depth-resolved profiles. \\[1ex]
\textbf{Transmission.} Corrected transmitted-energy accumulation.&
Transmission-energy summaries. \\[1ex]
\textbf{Recoils.} Explicitly initialized recoil counters. &
Recoil and sputtering summaries. \\[1ex]
\textbf{Layer handling.} Corrected selected layer-index updates. &
Layer-resolved quantities. \\
\bottomrule
\end{tabular}
\end{table*}

The historical implementation also assumed no more than five elements in a layer. The relevant arrays, loops, input checks, and output routines were generalized to a named maximum element count. This extension enables the simulation of target materials with more complex chemical compositions without altering the binary-collision transport model.

\subsection{Runtime-selectable RNG interface}

The historical \ranlux\ implementation~\cite{James1994CPC} was retained as the default and reference backend. Two optional generators, \pcg~\cite{ONeill2014} and \xoshiro~\cite{Blackman2021ATMS}, were placed behind the same uniform RNG interface and can be selected through an optional command-line argument. The existing seed fields in the input deck are used to initialize each backend, so no change to the input-file format is required.

Repeated calculations using the same input deck, RNG backend, and executable reproduce the same random sequence and simulation output. However, the sequences produced by different backends are intentionally different and exact agreement of individual histories is neither expected nor required. Consequently, the \ranlux\ path is used for exact regression testing, while the alternative backends are compared statistically.

\section{Validation methodology}

\subsection{Regression-testing strategy}

Changes were divided into three classes. First, comments, naming changes, routine extraction, and other structural refactoring were required to reproduce the output from the original code with the \ranlux\ RNG exactly, excluding timestamps and elapsed-time fields. Second, corrections to identified bookkeeping defects were allowed to change only the corresponding affected quantities. Third, RNG changes were required to be reproducible within each backend and statistically compatible with the reference results over the test suite.

After each behavior-preserving step, representative input decks were executed with \ranlux\ and the generated files were compared with the preceding reference version. Any difference was classified as a nondeterministic timing field, an intended correction, or an unintended numerical change before further modifications were accepted.

\subsection{Representative test suite}

The test suite was selected to exercise the code paths most relevant to \LEmSR\ and \bNMR. It contains single-layer stopping targets, multilayer structures, thin foils, targets with more than five elements per layer, and both $\mu^+$ and $^{8}$Li projectiles. Thin targets are particularly useful stress tests because small changes in individual trajectories can visibly change the balance among implantation, backscattering, and transmission. The complete set of tests is listed in Table~\ref{tab:test-cases}.

\begin{table*}[t]
\caption{Representative test cases used for regression, statistical comparison, and runtime measurements. $N$ is the number of incident projectiles at each energy. Layer thicknesses are given in \AA\ and are ordered from the incident side to the substrate side.}
\label{tab:test-cases}
\centering
\small
\begin{tabular}{lcrll}
\toprule
Test case & Projectile & $N$ & \shortstack{Energy scan\\(keV)} & \parbox[t]{0.43\textwidth}{Layer stack} \\
\midrule
6-element multilayer & $\mu^+$ & 50000 & 1--29, step 2 & \parbox[t]{0.43\textwidth}{AlCoCrFeNiTi(400) / SiO$_2$(400) / Al$_2$O$_3$(300) / Si(10000)} \\[0.5ex]
7-layer stack & $\mu^+$ & 50000 & 1--29, step 2 & \parbox[t]{0.43\textwidth}{SiO$_2$(400) / Al$_2$O$_3$(300) / TiN(300) / W(100) / TiN(200) / Al$_2$O$_3$(200) / Si(10000)} \\[0.5ex]
SiO$_2$/Al$_2$O$_3$/TiN/Si & $\mu^+$ & 50000 & 1--25, step 2 & \parbox[t]{0.43\textwidth}{SiO$_2$(400) / Al$_2$O$_3$(300) / TiN(300) / Si(10000)} \\[0.5ex]
Thin Au/Cr/Si & $\mu^+$ & 100000 & 1--19, step 2 & \parbox[t]{0.43\textwidth}{Au(100) / Cr(50) / Si(10000)} \\
Au/Cr/Si & $\mu^+$ & 50000 & 1--19, step 2 & \parbox[t]{0.43\textwidth}{Au(300) / Cr(300) / Si(10000)} \\
AlCoCrFeNiTi & $\mu^+$ & 50000 & 1--19, step 2 & \parbox[t]{0.43\textwidth}{AlCoCrFeNiTi(10000)} \\
SiO$_2$/Si & $\mu^+$ & 50000 & 1--19, step 2 & \parbox[t]{0.43\textwidth}{SiO$_2$(500) / Si(10000)} \\
SiO$_2$ & $\mu^+$ & 50000 & 1--19, step 2 & \parbox[t]{0.43\textwidth}{SiO$_2$(10000)} \\
Si & $\mu^+$ & 50000 & 1--19, step 2 & \parbox[t]{0.43\textwidth}{Si(10000)} \\
Thin C foil & $\mu^+$ & 100000 & 1--10, step 1 & \parbox[t]{0.43\textwidth}{C(100)} \\
Thin C/Au/C & $\mu^+$ & 100000 & 1--19, step 2 & \parbox[t]{0.43\textwidth}{C(100) / Au(300) / C(100)} \\
$^{8}$Li in Si & $^{8}$Li & 50000 & 1--19, step 2 & \parbox[t]{0.43\textwidth}{Si(10000)} \\
$^{8}$Li in thin C/Au/C & $^{8}$Li & 100000 & 1--19, step 2 & \parbox[t]{0.43\textwidth}{C(100) / Au(300) / C(100)} \\
\bottomrule
\end{tabular}
\end{table*}

The 13 target configurations contain 143 energy--configuration points. Each point was evaluated once with each of the three backends, giving 429 simulations. All backends used the same input deck and physical-model options at a given point.

\subsection{Integral outcome fractions}

For $N$ incident projectiles and $n_i$ events in outcome channel $i$, the estimated fraction is
\begin{equation}
 f_i=\frac{n_i}{N}.
\end{equation}
The implanted, backscattered, and transmitted fractions from a fast backend were compared with the corresponding \ranlux\ fraction using Newcombe's hybrid-score interval for the difference between two independent proportions~\cite{Newcombe1998,AgrestiCaffo2000}. For backends 1 and 2, the lower and upper limits are
\begin{align}
L_i={}&(f_{i,1}-f_{i,2})
 -\sqrt{(f_{i,1}-l_{i,1})^2+(u_{i,2}-f_{i,2})^2},\\
U_i={}&(f_{i,1}-f_{i,2})
 +\sqrt{(u_{i,1}-f_{i,1})^2+(f_{i,2}-l_{i,2})^2},
\end{align}
where the Wilson bounds for backend $k$ are
\begin{align}
l_{i,k}&=\frac{2Nf_{i,k}+z^2-z\sqrt{z^2+4Nf_{i,k}(1-f_{i,k})}}
 {2(N+z^2)},\\
u_{i,k}&=\frac{2Nf_{i,k}+z^2+z\sqrt{z^2+4Nf_{i,k}(1-f_{i,k})}}
 {2(N+z^2)},
\end{align}
with $z=1.96$ for a 95\% confidence level. An interval containing zero is compatible with no difference at this confidence level. An interval that excludes zero flags an individual comparison for inspection; however, this alone does not establish a systematic RNG-dependent bias across the test suite.

\subsection{Range and profile-shape comparisons}

Mean implantation depth and straggling were compared using absolute differences and parity plots. The output files do not contain all event-level information needed to reconstruct an uncertainty for every reported mean, so these quantities were treated as consistency diagnostics rather than formal equivalence tests. Particular caution is required when the number of implanted projectiles is small, because the reported mean range is then conditional on a sparse subset of the incident histories.

Depth-resolved implantation profiles, $F_i(z)$ where $i$ denotes the different RNGs, were compared both visually and through the maximum distance between their normalized cumulative profiles,
\begin{equation}
 D=\sup_z\left|F_1(z)-F_2(z)\right|,
 \label{eq:profile-distance}
\end{equation}
which is the statistic underlying the two-sample Kolmogorov--Smirnov (K--S) test~\cite{Kolmogorov1933,Smirnov1939}. Here, $D$ was used as a descriptive, dimensionless profile-shape distance. A formal K--S hypothesis test was not assigned because only binned profiles and one realization per backend were available, and the effective sample size is the number of implanted particles rather than the number of incident projectiles. The analysis therefore focuses on the magnitude and distribution of $D$, together with the peak position, width, and tail of representative profiles.

\subsection{Runtime measurements}

The runtime, $t$, was taken from the simulation-time field reported by the program. The speedup of backend $r$ relative to \ranlux\ is
\begin{equation}
 S_r=\frac{t_{\mathrm{RANLUX}}}{t_r}.
\end{equation}
All calculations used the same input decks, physical-model settings, and numbers of incident particles. Because wall-clock measurements can fluctuate with operating-system scheduling and file-system activity, individual timings are treated as practical indicators. The summed time over the complete suite is the primary performance metric.

\section{Results}

\subsection{Regression and functional tests}

The behavior-preserving refactoring retained the reference \ranlux\ execution path: deterministic outputs and Monte Carlo outputs generated from the same \ranlux\ sequence agreed with the preceding reference version after excluding timestamps and runtime fields. Differences caused by the localized corrections were confined to the quantities associated with the corrected code paths. The generalized target description also completed the six-element and seven-layer tests in Table~\ref{tab:test-cases}, demonstrating that the former element and layer assumptions no longer restrict these input decks.

\subsection{Runtime performance}

\begin{figure*}[t]
\centering
\safeincludegraphics[width=0.72\linewidth]{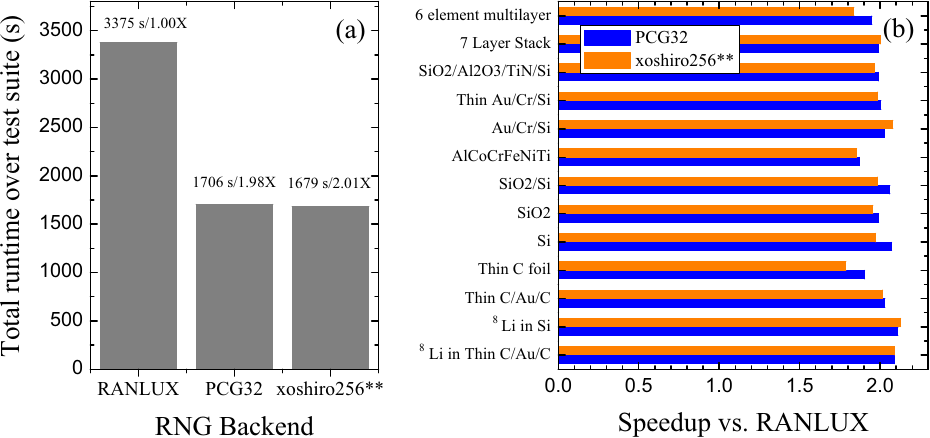}
\caption{(a) Total reported simulation time summed over the full test suite. (b) Runtime speedup relative to \ranlux\ for each test configuration. Both fast backends provide a systematic improvement across single-layer, multilayer, thin-film, $\mu^+$, and $^{8}$Li calculations.}
\label{fig:runtime}
\end{figure*}

Profiling of a representative 50000-projectile RANLUX calculation showed that the random-number-generation call path accounted for approximately 58.5\% of the profiled execution time. The calculation made $1.81 \times 10^8$ calls to the RANLUX routine, corresponding to approximately $3.6\times10^3$ RNG calls per incident projectile. In contrast, the extracted output-writing routines accounted for only about 0.6\% of the profiled runtime. Therefore, random-number generation represented a clear target for performance optimization.

Summed over all 143 points, the measured runtime was 3375~s for \ranlux, 1706~s for \pcg, and 1679~s for \xoshiro. The corresponding speedups were 1.98 and 2.01, respectively, as shown in Fig.~\ref{fig:runtime}. The per-configuration speedup ranged from 1.79 to 2.13 for \pcg\ and from 1.87 to 2.11 for \xoshiro. Smaller improvements occurred mainly for short calculations, for which fixed overhead and timing granularity represent a larger fraction of the total.

\subsection{Integral observables and range statistics}

\begin{figure*}[t]
\centering
\safeincludegraphics[width=\linewidth]{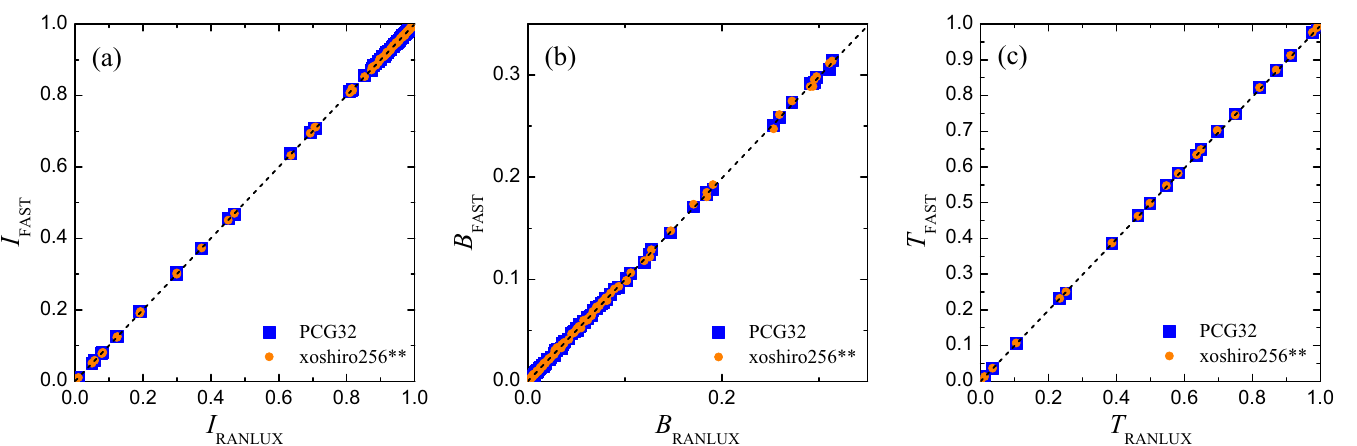}
\caption{(a) Implanted, (b) backscattered, and (c) transmitted fractions obtained with \pcg\ and \xoshiro\ plotted against the \ranlux\ reference. The dashed diagonal denotes exact agreement.}
\label{fig:fractions}
\end{figure*}

In Fig.~\ref{fig:fractions} we plot the implanted ($I$), backscattered ($B$), and transmitted ($T$) fractions of projectiles calculated using the \pcg\ and \xoshiro\ RNGs as a function of the corresponding values calculated from \ranlux. We find that these parity plots cluster closely around the line of equality. The mean absolute difference in $I$, $B$ and $T$ for \pcg\ and \xoshiro\ are summarized in Table~\ref{tab:observable-summary}. The 95th percentiles of the absolute differences in $I$, $B$ and $T$ ranged between $2.6\times10^{-3}$ and $3.2\times10^{-3}$ across the two fast RNGs.
\begin{table}
\caption{Mean absolute differences, $\langle|\Delta f|\rangle$, from RANLUX over the complete test suite. $I$, $B$, and $T$ denote implanted,
backscattered, and transmitted fractions, respectively. $R$ and $\sigma_R$ denote the implantation range/depth and straggling calculated from the implantation profiles, respectively.}
\label{tab:observable-summary}
\begin{ruledtabular}
\begin{tabular}{lccccc}
& \multicolumn{3}{c}{$10^{4} \times \langle|\Delta f|\rangle$}
& \multicolumn{2}{c}{$\langle|\Delta x|\rangle$ (nm)} \\
Backend & $I$ & $B$ & $T$ & $R$ & $\sigma_R$ \\
\hline
PCG32        & 9.4 & 8.5 & 1.8 & 0.092 & 0.076 \\
xoshiro256** & 8.6 & 8.6 & 1.8 & 0.128 & 0.089 \\
\end{tabular}
\end{ruledtabular}
\end{table}

\begin{figure*}[htb]
\centering
\safeincludegraphics[width=0.68\linewidth]{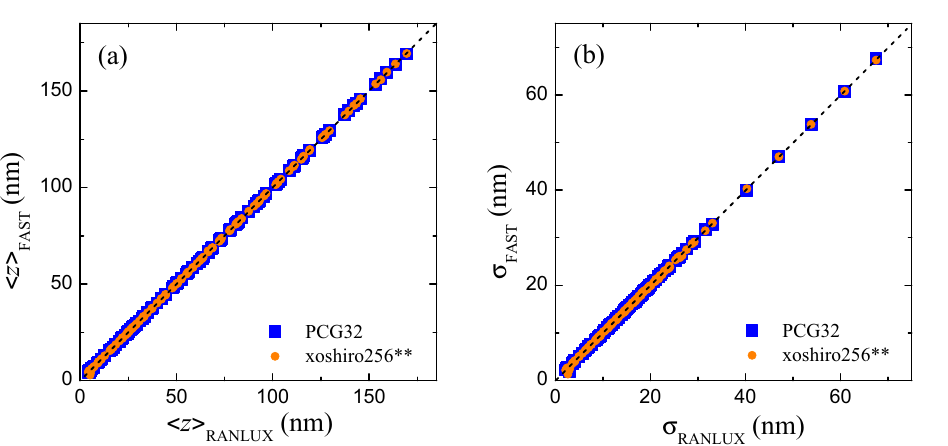}
\caption{(a) Mean implantation depth and (b) range straggling obtained with \pcg\ and \xoshiro\ plotted against the \ranlux\ reference. The dashed diagonal denotes exact agreement.}
\label{fig:range-parity}
\end{figure*}

The mean absolute range differences were 0.092~nm for \pcg\ and 0.128~nm for \xoshiro, while the mean absolute differences in straggling were 0.076 and 0.089~nm, respectively (Table~\ref{tab:observable-summary}). The largest individual range differences were 1.10~nm for \pcg\ and 2.49~nm for \xoshiro. These occurred in broad or sparsely populated profiles and are visible as isolated deviations in Fig.~\ref{fig:range-parity}; they do not form a coherent offset from the diagonal.

\subsection{Confidence-interval comparison of outcome fractions}

\begin{figure}[htb]
\centering
\safeincludegraphics[width=\linewidth]{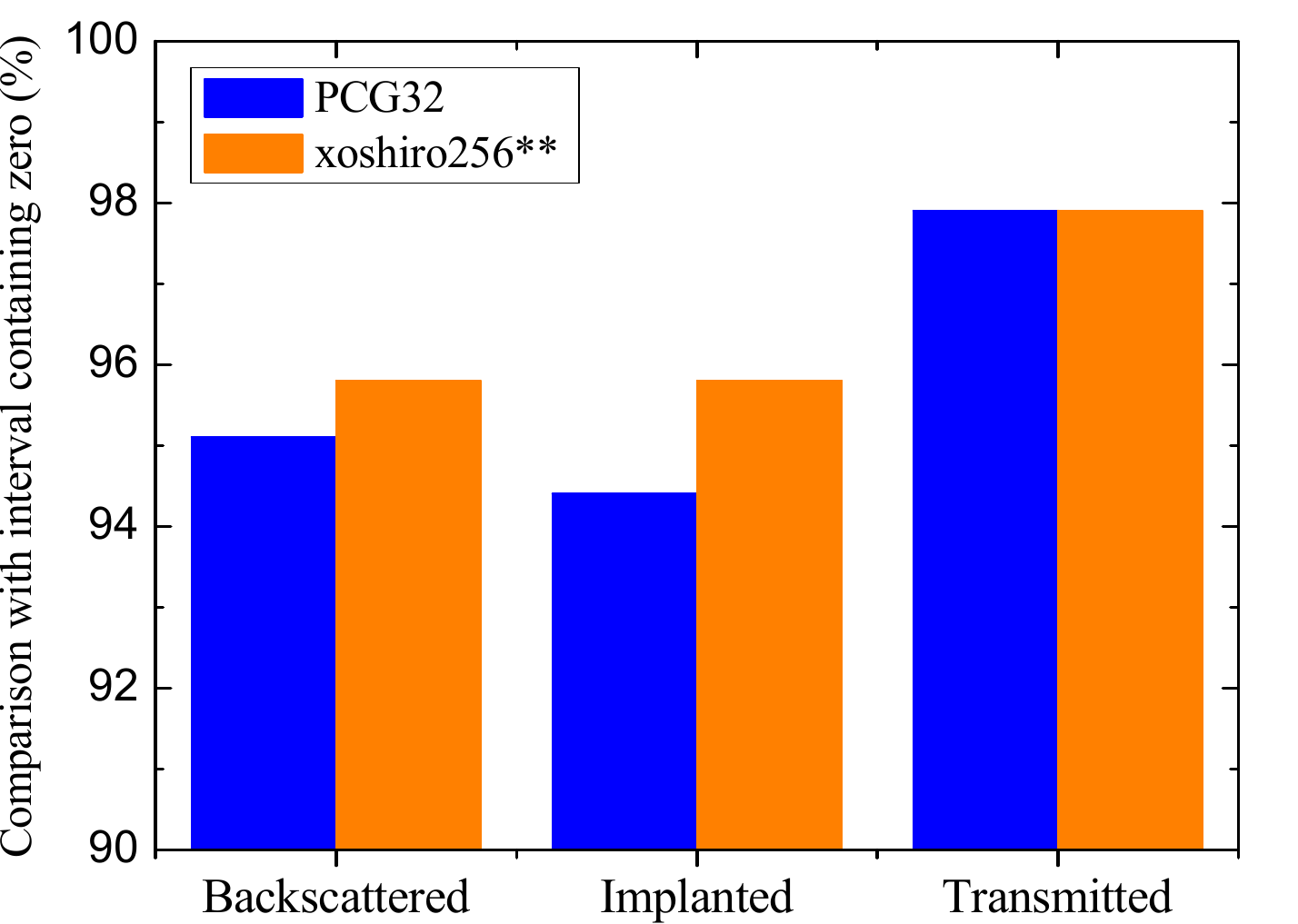}
\caption{Fraction of the 143 energy--configuration comparisons for which the 95\% Newcombe interval for the difference from \ranlux\ contains zero. The corresponding numbers outside the interval are 7, 8, and 3 for \pcg\ and 6, 6, and 3 for \xoshiro\ for the backscattered, implanted, and transmitted channels, respectively.}
\label{fig:newcombe}
\end{figure}

Figure~\ref{fig:newcombe} summarizes the Newcombe intervals for the differences between each fast RNG and \ranlux. If an interval contains zero, the observed difference is statistically compatible with no difference between the two RNGs at the 95\% confidence level. For \pcg, this was the case for 136 of 143 backscattered comparisons (95.1\%), 135 of 143 implanted comparisons (94.4\%), and 140 of 143 transmitted comparisons (97.9\%). The corresponding numbers for \xoshiro\ were 137 of 143 (95.8\%), 137 of 143 (95.8\%), and 140 of 143 (97.9\%). Thus, approximately 94-98\% of the individual comparisons were compatible with zero difference at 95\% confidence level, with no outcome channel or backend showing a systematically elevated interval-exclusion rate.

The comparisons outside the interval were distributed among several configurations. They occurred most often when an outcome channel contained relatively few events or when thin-target transport produced a rapidly changing balance among implantation, backscattering, and transmission. In such cases, a difference of only a small number of histories can be prominent relative to the rare channel.

\subsection{Depth-resolved profiles and sparse thin-target cases}

\begin{figure*}[t]
\centering
\safeincludegraphics[width=0.8\linewidth]{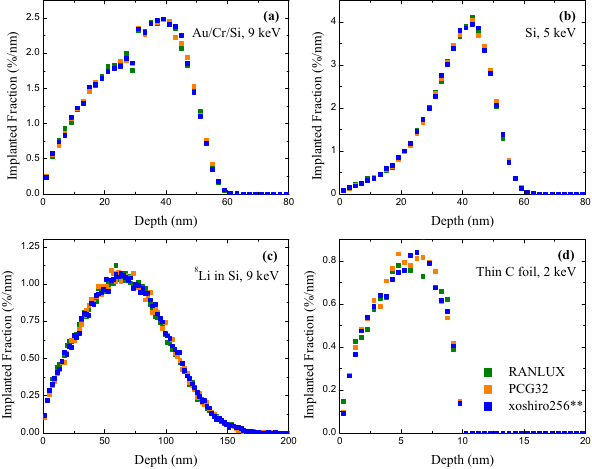}
\caption{Representative implantation profiles for (a) $\mu^+$ in Au (30~nm) / Cr (30~nm) / Si at 9~keV, (b) $\mu^+$ in Si at 5~keV, (c) $^{8}$Li in Si at 9~keV, and (d) $\mu^+$ in a thin carbon foil at 2~keV. The three RNG backends reproduce the same principal profile features.}
\label{fig:profiles}
\end{figure*}

The mean cumulative-profile distance defined in Eq.~\eqref{eq:profile-distance} was 0.014 for \pcg\ and 0.017 for \xoshiro. The corresponding 95th percentiles were 0.020 and 0.038, respectively. Representative profiles are shown in Fig.~\ref{fig:profiles}. For well-populated thick targets and multilayers, the peak position, width, and tail are closely reproduced by all three RNGs.

The larger profile distances occur primarily in sparse thin-foil conditions. At energies for which most projectiles transmit through the foil, the implanted profile is constructed from only a small fraction of the incident histories. Normalizing this sparse profile amplifies the effect of individual trajectories. The conditional mean range can consequently fluctuate strongly even while the dominant transmission fraction remains stable. Such points are retained as stress tests but should not be interpreted in the same manner as well-populated stopping profiles. 

\subsection{RNG backend selection}

Both fast RNG backends provide essentially the same end-to-end performance gain and comparable agreement with the reference observables. The present results, therefore, do not establish a physics-based preference between \pcg\ and \xoshiro. \ranlux\ remains the appropriate default when continuity with historical calculations or exact regression of the legacy sequence is required. For routine parameter scans, either fast backend can be selected without altering the input deck. A project may choose one as its documented default to simplify reproducibility across calculations performed with the same documented software release.

\section{Discussion and limitations}

The validation demonstrates that the modernization preserves the established \ranlux\ path for behavior-preserving changes and that the two alternative RNGs do not produce a systematic shift in the tested observables. It also demonstrates a practical speed improvement by a factor of two for the defined workstation test suite. These conclusions are deliberately limited to the simulated projectiles, energies, target geometries, and model options listed in Table~\ref{tab:test-cases}.

The comparison is an internal validation of software behavior rather than an experimental validation of the TRIM.SP transport model. Agreement among RNG backends cannot establish the absolute accuracy of stopping powers, scattering potentials, sputtering models, or predicted implantation profiles. Those questions require comparison with experimental data (see, e.g., Ref.~\cite{McFadden2026NIaMiPRSBBIwMaAa}) or independent transport calculations (see, e.g., Ref.~\cite{Saquib2012PP}).

A further limitation is that each energy-configuration point was evaluated once per backend. The Newcombe analysis is appropriate for the integral binomial outcome counts, but the mean range and profile-shape comparisons do not provide formal equivalence bounds. A future targeted study with repeated independent seeds would quantify within-backend run-to-run variation and permit a direct comparison with between-backend variation. Such repetitions are most useful for a smaller set of representative thick, multilayer, thin-foil, and $^{8}$Li cases rather than for the entire scan matrix.

Finally, the transport kernel remains largely legacy Fortran. Continued decomposition into smaller testable units would improve maintainability, but each step must retain the present distinction between behavior-preserving refactoring, intentional defect correction, and changes to the scientific model. Extension to substantially higher implantation energies should likewise be preceded by a dedicated review of stopping, depth-binning, and target-thickness assumptions.

\section{Conclusions}

A local generalized multilayer extension of TRIM.SP has been reorganized into the \trimspnl\ transport engine and coupled to the GUI-based \workbench. The modernization separated the input and output infrastructure from the transport flow, introduced explicit build modes and documentation, corrected localized bookkeeping defects, removed the former five element per layer restriction, and added runtime selection of RNG among \ranlux, \pcg, and \xoshiro.

Exact regression testing retained the historical \ranlux\ path for behavior-preserving changes. Across 13 representative configurations and 429 simulations, the two fast RNG backends reduced the summed runtime by approximately a factor of two. Integral outcome fractions, range quantities, and representative depth profiles showed no systematic backend-dependent trend; the largest visible deviations were associated with sparse implanted populations in thin targets.

The resulting design preserves \ranlux\ for historical continuity while enabling substantially faster GUI-driven scans with either modern backends. The open-source release provides a practical basis for further testing and incremental modernization without replacing the established binary-collision transport model.

\section{Acknowledgments}
This research was supported by the NCCR Muoniverse, a National Centre
of Competence in Research, funded by the Swiss National Science
Foundation (SNSF; grant No.~51NF-0\_229254), and by the SNSF project
``Hybrid molecule/inorganic quantum material interfaces''
(grant No.~\href{https://data.snf.ch/grants/grant/238694}{2025-07436}).

\section*{Code availability}
The source code, build files, and GUI workflow are available from the public PSI Gitea repository \href{https://gitea.psi.ch/LMU/TRIMSP}{TRIMSP-NL Workbench}. The version described in this work corresponds to release \texttt{v1.4.0} (commit \texttt{19b9eacf53c1}).

\section*{Declaration of interest}

The authors declare that they have no known competing financial
interests or personal relationships that could have appeared to
influence the work reported in this paper.

\section*{Declaration of generative AI and AI-assisted technologies
in the manuscript preparation process}
During the preparation of this work, the authors used ChatGPT (OpenAI) to assist with manuscript organization and language editing. The Authors reviewed and verified all content and take full responsibility for the publication.

\bibliography{references}

\end{document}